\documentclass[notitlepage,onecolumn]{revtex4-1}
\usepackage{amsmath,amssymb,amsthm,mathtools}
\usepackage{euscript,mathrsfs}
\usepackage{graphicx}
\usepackage{wrapfig}
\usepackage[dvipsnames]{xcolor}
\usepackage{wasysym}
\usepackage{verbatim}
\usepackage{empheq}
\usepackage{adjustbox}
\usepackage{xcolor}
\usepackage{mathrsfs} 
\usepackage{hyperref}
\definecolor{urlcolor}{HTML}{990000}
\definecolor{linkcolor}{HTML}{005F5F} 
\hypersetup{pdfstartview=FitH,  linkcolor=linkcolor,urlcolor=urlcolor, colorlinks=true,citecolor=blue}
\makeatletter
\def\@ssect@ltx#1#2#3#4#5#6[#7]#8{%
  \def\H@svsec{\phantomsection}%
  \@tempskipa #5\relax
  \@ifdim{\@tempskipa>\z@}{%
    \begingroup
      \interlinepenalty \@M
      #6{%
       \@ifundefined{@hangfroms@#1}{\@hang@froms}{\csname @hangfroms@#1\endcsname}%
       {\hskip#3\relax\H@svsec}{#8}%
      }%
      \@@par
    \endgroup
    \@ifundefined{#1smark}{\@gobble}{\csname #1smark\endcsname}{#7}%
  }{%
    \def\@svsechd{%
      #6{%
       \@ifundefined{@runin@tos@#1}{\@runin@tos}{\csname @runin@tos@#1\endcsname}%
       {\hskip#3\relax\H@svsec}{#8}%
      }%
      \@ifundefined{#1smark}{\@gobble}{\csname #1smark\endcsname}{#7}%
      \addcontentsline{toc}{#1}{\protect\numberline{}#8}%
    }%
  }%
  \@xsect{#5}%
}%
\makeatother

\renewcommand{\phi}{\varphi}

\numberwithin{equation}{section}

\begin{document}

\title{Synchronization on star graph with noise}
\author{Artem Alexandrov}
\affiliation{Moscow Institute of Physics and Technology, Dolgoprudny 141700, Russia}

\begin{abstract}
We investigate synchronization in the Kuramoto model with noise on a star graph. By revising the case of a complete graph, we propose a closed form of self-consistency equation for the conventional order parameter and generalize it for a star graph. Using the obtained self-consistency equation, we demonstrate that there is a crossover between the abrupt synchronization at small noise and the continuous phase transition for quite large noise. We probe this crossover numerically and analytically.
\end{abstract}

\date{\today}
        
\maketitle

\tableofcontents

\section{Introduction}

Synchronization phenomena occur in different artificial and natural systems. There are several common models that capture the synchronization transition. One of the most famous of them is the Kuramoto model \cite{Kuramoto1984}.

Kuramoto model with all-to-all couplings (which can be considered as the model on a complete graph) is studied intensively. It seems that the current understanding of the complete graph case is quite comprehensive. In the early papers it was shown that synchronization transition is the second order phase transition with respect to the coupling constant \cite{Kuramoto1984,SakaguchiKuramoto1986}. The case of identical oscillator frequencies especially attracts lot of attention because the model on a complete graph exhibits low-dimensional dynamics, which can be observed by applying Watanabe-Strogatz (W-S) ansatz \cite{Watanabe1993}. This appealing and quite unexpected low-dimensional dynamics emerges due to existence of integrals of motion. These integrals of motion were initially proposed in \cite{Watanabe1993,Watanabe1994}. The rigorous explanation how exactly the integrals of motion appear is given in \cite{Marvel2009}, where the direct connection between W-S ansatz and M\"{o}bius transformation is established. Next, the connection between Ott-Antonsen (O-A) \cite{Ott2008} ansatz and W-S ansatz was discussed in \cite{Pikovsky2011}. The emergent hyperbolic geometry picture for the Kuramoto model with identical oscillators is proposed in \cite{Chen2017}. In addition, the case of coupled complete graphs is considered in \cite{Pikovsky2008}, where the possibility of the chimera state \cite{Abrams2008} is discussed and several insights about thermodynamic limit are proposed.

The case of more general graph topology is not so deeply explored. For instance, there are only a few examples of graphs which allow the analytic study of the synchronization transition. One of these examples is the so-called star graph and Cartesian products of graphs. The star graph was considered numerically and analytically in \cite{Vlasov2015,Boccaletti2019}. In papers \cite{Wang2017,Chen2019} the authors have succeeded in the consideration of the Kuramoto model on Cartesian graph products. The case of chain and ring graphs are partially covered in \cite{Ochab2009,Roy2012}. An attempt to consider more general star-like graphs is discussed in \cite{Alexandrov2021}. One of the most interesting facts concerning more general topology is that the synchronization transition can be the first order transition (abrupt synchronization) if the oscillator frequencies correlate with node degrees \cite{Arenas2011ExplosiveSync,Coutinho2013,Arenas2019Review}.

At least two modifications of Kuramoto model are widely used: the so-called second order Kuramoto model \cite{Tanaka1997} (also called Kuramoto model with inertia), which contains the second time derivative of oscillator phase and the Kuramoto model with noise \cite{Sakaguchi1988} (so-called Kuramoto-Sakaguchi model). The second order Kuramoto model with all-to-all coupling exhibits hysteresis and attracts lot of research interest in recent years \cite{Olmi2014}. Also, the model with second time derivative and noise is discussed in \cite{Acebron2000,Gupta2014}. The case of all-to-all coupling with noise was considered in pioneering work by Sakaguchi \cite{Sakaguchi1988}. It was shown that noise affects the synchronization transition. The effect of noise can be captured in an alternative approach based on linear stability analysis in \cite{Mirollo1990,Strogatz1991}. Further, in the papers \cite{Bag2007,Tonjes2010} more general cases of noise are considered (colored noise \& Ornstein-Ulenbeck process).

The noise is a common actor in biological systems. A lot of research at this landscape relates to biological neural networks. The impact of noise on these networks is discussed in \cite{Tabareau2010}, where the connections between neuron synchronization and intrinsic random perturbations are considered. The synchronization by noise is investigated in the rigorous mathematical way in \cite{Flandoli2017} and has the interesting addition by consideration of the noise-induced synchronization in Kuramoto model on 2D lattice \cite{Sarkar2020}. Several aspects of the interplay between synchronization and noise is discussed in the Kuramoto model framework for KKI-18 large human connectome graph \cite{Odor2021}.

In this note we investigate the synchronization transition in Kuramoto model on a star graph with noise. The rest of paper is organized as follows. In the \autoref{sec:CompleteGraph}, we start our research by considering the well-known case of complete graph and propose a closed form of a self-consistency equation for the conventional order parameter. We test validity of our derivations by rederiving the critical coupling constant for complete graph case. 

Next, in the \autoref{sec:StarGraphNoise}, we try to generalize our approach on the case of star graph. We demonstrate that the appeared self-consistency equation is much more complicated in compare with complete graph case. To avoid these complications, we apply the O-A ansatz (we discuss its applicability with noise in the \autoref{sec:CompleteGraph}). The obtained equation coincides with already known result for $D=0$ \cite{Vlasov2015}. Our treatment indicates that the synchronization transition becomes continuous for large enough values of noise, whereas the case of small noise exhibits the abrupt synchronization, i.e. the transition is discontinuous.

We finalize our investigation by the \autoref{sec:Simulations}, where we provide results of simulations that confirm our theoretical propositions and speculations. The \autoref{sec:Conclusion} summarizes presented results and contains discussion of further plans. In the appendices, \autoref{sec:TechnicalDetails} \& \autoref{sec:SimulationDetails}, we provide the calculation of essential quantities appeared in the main text and all the details about simulations.

\section{Complete graph}
\label{sec:CompleteGraph}

\subsection{Self-consistency equation}
Here we revise some aspects of original paper \cite{Sakaguchi1988}. Consider the Kuramoto-Sakaguchi model on the complete graph,
\begin{equation}\label{cg-eom}
    \dot{\theta}_i = \omega_i + \frac{\lambda}{N}\sum_{j=1}^{N}\sin(\theta_j-\theta_i) + \eta_i,
\end{equation}
where $\eta_i=\eta_i(t)$ is the white noise, so its time average is zero, $\langle\eta_i(t)\rangle=0$, whereas two-time correlator is not zero, $\langle\eta_i(t)\eta_j(t')\rangle = 2D\delta_{ij}\delta(t-t')$. The variable $\theta_i$ represents phase oscillator, $\theta_i\in[-\pi,\pi)$ and $\lambda>0$ is the coupling constant, $N$ denotes the number of oscillators. The conventional order parameter is defined by
\begin{equation}\label{op-definition}
    r(t) = \frac{1}{N}\sum_{j=1}^{N}e^{i\theta_j(t)}
\end{equation}
and in the synchronized state it does not depend on time. Substituting this expression into the equations of motion~\eqref{cg-eom}, we find
\begin{equation}\label{cg-eom-op}
    \dot{\theta}_i = \omega_i+\lambda |r|\sin(\arg r-\theta_i)+\eta_i.
\end{equation}
It is the set of Langevin equations, which in the limit of large number of nodes, $N\rightarrow\infty$, corresponds to the Fokker-Planck equation for the distribution $\rho=\rho(\theta,\omega,t)$ \cite{Lancellotti2005},
\begin{equation}\label{fp-equation}
    \frac{\partial\rho}{\partial t}=-\frac{\partial}{\partial\theta}\left\{\left(\omega+\lambda|r|\sin(\arg r-\theta)\right)\rho(\theta,\omega,t)\right\}+D\frac{\partial^2\rho}{\partial t^2},
\end{equation}
where the distribution function has the following properties,
\begin{equation}\label{fp-sol-properties}
    \int_{-\pi}^{+\pi}d\theta\,\rho(\theta,\omega,t)=1,\quad \rho(\theta+2\pi,\omega,t)=\rho(\theta,\omega,t).
\end{equation}
In continuum limit, the order parameter becomes
\begin{equation}\label{op-fp}
    r(t) = \int_{-\infty}^{+\infty}d\omega\,g(\omega)\int_{-\pi}^{+\pi}d\theta\,\rho(\theta,\omega,t)e^{i\theta},
\end{equation}
where $g(\omega)$ is a given frequency distribution. The synchronized state corresponds to a non-trivial stationary solution of this Fokker-Planck equation. The ansatz for such solution is given by
\begin{equation}\label{fp-sol-ansatz}
    \rho(\theta,\omega)=C_1e^{f(\theta)}\int_{C_2}^{\theta}d\phi\,e^{-f(\phi)},\quad f(\phi)=\frac{\omega\phi+\lambda |r|\cos(\arg r-\phi)}{D}.
\end{equation}
The quantities $C_1$ and $C_2$ do not depend on time and can be found with help of eq.~\eqref{fp-sol-properties}. To determine the values of $C_1$ and $C_2$, we follow \cite{Stratonovich1967} (ch. 9, sec. 2, pp. 236-239). The resulting expression for the stationary distribution is
\begin{equation}\label{fp-stationary-sol}
    \rho(\theta,\omega) = \frac{e^{\pi\omega/D}e^{\{\omega\theta+\lambda |r|\cos(\arg r-\theta)\}/D}}{4\pi^2I_{i\omega/D}(\lambda |r|/D)I_{-i\omega/D}(\lambda |r|/D)}\int_{\theta}^{2\pi+\theta}d\phi\,e^{-(\omega\phi+\lambda|r|\cos\{\arg r-\phi\})/D}.
\end{equation}
This solution corresponds to the case of non-identical oscillators, i.e. the eigenfrequencies in eq.~\eqref{cg-eom-op} are different. In case of identical frequencies, there is the great simplification. With identical frequencies, $\forall i$, $\omega_i\equiv\omega$ one can use rotating reference frame in~\eqref{cg-eom-op}, which effectively corresponds to the setting $\omega=0$ in~\eqref{fp-stationary-sol}. In such case, the distribution function $\rho(\theta,\omega)$ becomes
\begin{equation}
    \rho(\theta,0)=\frac{1}{2\pi I_0(\lambda|r|/D)}\exp\left\{\frac{\lambda|r|}{D}\cos(\arg r-\theta)\right\},
\end{equation}
which is nothing more than von Mises distribution. These computations may be a little bit confusing but this misunderstanding should disappear if one keeps in mind that the frequency distribution $g(\omega)$ arises a posteriori in the treatment of Kuramoto-Sakaguchi model \cite{Lancellotti2005}. Having found the stationary solution of Fokker-Planck equation, we can easily obtain the self-consistency equation. We substitute eq.~\eqref{fp-stationary-sol} into the expression for order parameter~\eqref{op-fp} and perform integration over the phase $\theta$, which gives us the following integral form of self-consistency equation,
\begin{equation}\label{cg-self-consist}
    r = \int_{-\infty}^{+\infty}d\omega\,g(\omega)\frac{I_{1-i\omega/D}(\lambda|r|/D)}{I_{-i\omega/D}(\lambda|r|/D)}.
\end{equation}
The appeared ratio of Bessel functions contains imaginary and real part that depend on $\omega$. The real part is even with respect to $\omega$, whereas the imaginary part is odd in $\omega$. For symmetric frequency distributions, $g(\omega)=g(-\omega)$, we can represent the self-consistency as
\begin{equation}
    |r| = \frac{1}{2}\int_{-\infty}^{+\infty}d\omega\,g(\omega)\left[\frac{I_{1+i\omega/D}(\lambda|r|/D)}{I_{i\omega}(\lambda|r|/D)}+\frac{I_{1-i\omega/D}(\lambda|r|/D)}{I_{-i\omega}(\lambda|r|/D)}\right].
\end{equation}
At the vicinity of critical point, we can consider that $|r|$ is small. Expanding Bessel functions with respect to $|r|$, we obtain exactly the same expression as in Sakaguchi's paper,
\begin{equation}
    |r| = \frac{\lambda D|r|}{2}\int_{-\infty}^{+\infty}\frac{d\omega\,g(\omega)}{\omega^2+D^2}-\frac{\lambda^3|r|^3D}{4}\int_{-\infty}^{+\infty}\frac{d\omega\,g(\omega)(D^2-2\omega^2)}{(D^2+\omega^2)(4D^2+\omega)^2}+\mathcal{O}\left(|r|^4\right).
\end{equation}
The non-trivial solution $|r|\neq 0$ of self consistency equation exists if $\lambda>\lambda_c$, where the expression for $\lambda_c$ is given by
\begin{equation}\label{critical-coupling}
    \frac{1}{\lambda_c}=\frac{D}{2}\int_{-\infty}^{+\infty}\frac{d\omega\,g(\omega)}{\omega^2+D^2}.
\end{equation}
It is worth mentioning that in case of identical frequencies, the expression for order parameter is nothing more than first moment of von Mises distribution and the resulting self-consistency equation has the very simple form,
\begin{equation}\label{sc-eq-cg-ident}
    |r| = \frac{I_1(\lambda|r|/D)}{I_0(\lambda|r|/D)}
\end{equation}
and coincides with self-consistency for XY model and for HMF model with $D$ plays role of temperature (the critical constant in this case is simply $\lambda_c=2D$). Finally, the expression for $\lambda_c$, eq.~\eqref{critical-coupling}, can be derived via linear stability analysis of incoherent state, which was shown in the work \cite{Strogatz1991}.

\subsection{Ott-Antonsen ansatz in presence of noise}

In case of zero noise, the dynamics of Kuramoto model can be studied with help of O-A ansatz \cite{Ott2008}. This ansatz is applicable in continuum limit and based on the quite simple idea that distribution function $\rho=\rho(\theta,\omega,t)$ can be represented by Fourier series,
\begin{equation}
    \rho(\theta,\omega,t)=\frac{1}{2\pi}\sum_{m=-\infty}^{+\infty}\alpha_m(\omega,t)e^{im\theta},\quad \overline{\alpha}_m=\alpha_{-m},\,\alpha_m\in\mathbb{C}.
\end{equation}
Direct substitution of this Fourier series into an equation for $\rho$ gives an infinite hierarchy of differential equations for $\alpha_m$. Hopefully, the higher modes $\alpha_m$ can be represented via the very first mode $\alpha_1$ as $\alpha_m=(\alpha_1)^m$, which is called Ott-Antonsen ansatz.

Naive application of O-A ansatz to the Kuramoto-Sakaguchi model gives the following set of equations for $\alpha_m$,
\begin{equation}\label{o-a-noise}
    \dot{\alpha}_m = i\omega m\alpha_m + \frac{m\lambda r}{2}\alpha_{m-1}-\frac{\lambda m\overline{r}}{2}\alpha_{m+1}-m^2D\alpha_m,
\end{equation}
where the order parameter $r$ can be expressed via $\alpha_{-1}$,
\begin{equation}\label{o-a-op}
    r = \int_{-\infty}^{+\infty}d\omega\,g(\omega)\alpha_{-1}.
\end{equation}
It is easy to see that the substitution $\alpha_m=(\alpha_1)^m$ does not decouple modes due to the additional term with noise amplitude. However, at the vicinity of critical point, where the order parameter is small, one can still consider O-A ansatz. Setting $m=1$, we find
\begin{equation}\label{o-a-noise-truncated}
    \dot{\alpha} = i\omega \alpha + \frac{\lambda r}{2}-\frac{\lambda \overline{r}}{2}\alpha^{2}-D\alpha,
\end{equation}
where $\alpha_1\equiv \alpha$. Because we are interested in the critical point, we look for the fixed point of~\eqref{o-a-noise-truncated}. Substituting the fixed point into eq.~\eqref{o-a-op} we again find the self-consistency equation for $|r|$ but in slightly different form and the result for critical constant $\lambda_c$ coincides (as should be) with eq.~\eqref{critical-coupling}. It means that despite the presence of noise, we can analyze the equation for $\alpha_1$ in order to determine the critical point. It causes by the simple fact that the conventional order parameter contains only the first mode $\alpha_{-1}$ (in case of identical oscillators the order parameter is simply $r\equiv\alpha_{-1}$).

To finalize this subsection, let us draw attention to the following speculative idea. The obtained eq.~\eqref{o-a-noise-truncated} is nothing more than the Riccati equation and one can check that (after some algebraic manipulations) the ratio of Bessel functions from~\eqref{cg-self-consist} solves this equation. During this derivations, one should start to measure time in terms of inverse coupling constant, $\lambda^{-1}$. At first glance, it may seem strange, but it tempts us to say that something similar to renormalization group can be established in such model. We do not want to speculate about it, but simply highlight this possibility.

\section{Star graph with noise}
\label{sec:StarGraphNoise}

Armed with knowledge about closed form expression for the order parameter in case of a complete graph, we now consider a little bit more complicated case, a star graph. The analytical results for the Kuramoto model on a star graph were obtained in work \cite{Vlasov2015}.

Before we start our analysis of the Kuramoto model on a star graph with noise, we would like to revise several important (for further narration) moments of the mentioned original paper by Vlasov et al \cite{Vlasov2015}. In this papers, the authors considered the Kuramoto model on a star graph, which equations of motion are given by
\begin{equation}
    \dot{\Phi}= \beta\omega + \frac{\lambda\beta}{N}\sum_{j=1}^{N}\sin\left(\theta_j-\Phi\right);\quad 
    \dot{\theta}_j = \omega + \lambda\sin\left(\Phi-\theta_j\right),
\end{equation}
where $\Phi$ is the central node phase (hub phase) and $\theta_j$, $i\in\{1,N\}$ are phases of leaves, $\lambda>0$ is the coupling constant. All the leaves have identical eigenfrequencies $\omega$, whereas the hub has frequency $\beta\omega$, where $\beta>0$ is constant. We call the constant $\beta$ hub frequency enhancement factor. Introducing the phase difference, $\phi_j=\theta_j-\Phi$, one obtains the equation that allows to apply Watanabe-Strogatz ansatz. Then, in the thermodynamic limit, $N\rightarrow\infty$, the discontinuous phase transition occurs, which corresponds to abrupt synchronization. The critical points of such transition are given by
\begin{equation}\label{star-cc}
    \lambda_c^f=\frac{(\beta-1)\omega}{\sqrt{2\beta+1}},\quad \lambda_c^b=\frac{(\beta-1)\omega}{\beta+1}.
\end{equation}
The constant $\lambda_c^f$ corresponds to the evolution from nonsynchronized state with increasing of the coupling constant (forward evolution), the constant $\lambda_c^b$ corresponds to the synchronized state decay when the coupling constant decrease (backward evolution). So, the synchronization transition is the discontinuous (the first order) transition and the system exhibits hysteresis.

\subsection{Order parameter for star graph with noise}

Now let us consider the Kuramoto model on a star graph with noise, which equations of motion are
\begin{equation}
    \dot{\Phi}= \beta\omega + \frac{\lambda\beta}{N}\sum_{j=1}^{N}\sin\left(\theta_j-\Phi\right);\quad 
    \dot{\theta}_j = \omega + \lambda\sin\left(\Phi-\theta_j\right) + \eta_j,
\end{equation}
with $\eta_j$ is again white noise. We consider the Dirac delta function distribution of eigenfrequencies, i.e. all the leaves have frequency $\omega$ and the hub has frequency $\beta\omega$. We make this choice in order to compare our results with the case of zero noise. These equations of motion can be rewritten by introducing phase difference between leaves and hub, $\phi_j = \theta_j-\Phi$, which gives us
\begin{equation}\label{eom}
    \dot{\phi}_i = -(\beta-1)\omega-\lambda\sin\phi_i-\frac{\lambda\beta}{N}\sum_{j=1}^{N}\sin\phi_i+\eta_i.
\end{equation}
Following \cite{Vlasov2015}, we consider the order parameter,
\begin{equation}
    r = \frac{1}{N}\sum_{j=1}^{N}e^{i(\theta_j-\Phi)}=\frac{1}{N}\sum_{j=1}^{N}e^{i\phi_j}.
\end{equation}
In continuum limit, this expression becomes,
\begin{equation}
    r(t)=\int_{-\pi}^{+\pi}d\phi\,e^{i\phi}\rho(\phi,\,\omega,\,t),
\end{equation}
where $\rho(\phi,\,\omega,t)$ is the solution of the following Fokker-Planck equation,
\begin{equation}\label{F-P-star}
    \frac{\partial \rho(\phi,\omega,t)}{\partial t} = -\frac{\partial}{\partial\phi}\left\lbrace\left[-(\beta-1)\omega-\lambda\sin\phi -\lambda\beta\,\text{Im}\,r\right]\rho(\phi)\right\rbrace+D\frac{\partial^2\rho(\phi)}{\partial\phi^2},
\end{equation}
Let us introduce the following quantities
\begin{equation}
    A = -\frac{(\beta-1)\omega+\lambda \beta\,\text{Im}\,r}{D},\quad B = +\frac{\lambda}{D}.
\end{equation}
In terms of these quantities, we have actually the same equation as for the complete graph case. Therefore, we can immediately write down the stationary solution,
\begin{equation}
    \rho(\phi,\omega)=\frac{e^{A\phi+B\cos\phi}}{4\pi^2e^{-\pi A}I_{-iA}(B)I_{iA}(B)}\int_{\phi}^{\phi+2\pi}d\psi\,e^{-A\psi-B\cos\psi}
\end{equation}
and then can write down the self-consistency equation,
\begin{equation}\label{op-star-graph}
    r = \frac{I_{1-iA}(B)}{I_{-iA}(B)}.
\end{equation}
In last expression, one should consider real and imaginary part separately and then compute $|r|$. The complication (in compare with a complete graph case) arises from the presence of imaginary part of order parameter. Indeed, this ratio of modified Bessel functions is nor even neither odd in $\omega$, so the order parameter has both imaginary and real parts (in case of complete graph, one can safely set the imaginary part to zero in case of symmetric $g(\omega)$). The resulting expression~\eqref{op-star-graph} should be considered near critical point (which corresponds to small $r$). However, one more complication arises: in case of a complete graph, we deal with the simple expansion of modified Bessel function for small argument whereas for a star graph the order parameter is contained in the order of Bessel functions. One can try to analyze the eq.~\eqref{op-star-graph} numerically, but there a lot of parameters: the hub frequency enhancement factor $\beta$, the coupling constant $\lambda$, the noise amplitude $D$, the frequency $\omega$. In addition, we should care about both imaginary and real parts of the order parameter. In principle, we can realize this approach but it is labour-intensive. The second possible way to capture phase transition is to analyze stability of small perturbation near incoherent stare. Unfortunately, this method is also quite complicated and we do not provide our attempts to do it. Fortunately, as we have noticed earlier, the Ott-Antonsent ansatz captures the phase transition point for the system with noise.

\subsection{Critical points}

Representing the distribution function $\rho(\phi,\omega,t)$ via its Fourier series, we obtain the following set of equations for Fourier modes $\alpha_m$,
\begin{equation}\label{o-a-hierarchy-ansatz-star}
    \dot{\alpha}_m = -im(\beta-1)\omega\alpha_m + \frac{\lambda m}{2}\alpha_{m-1}-\frac{\lambda m}{2}\alpha_{m+1}-im\lambda\beta\,\mathrm{Im}\,r\alpha_m - m^2D\alpha_m.
\end{equation}
Setting $m=1$ and using the Ott-Antonsen ansatz, we find
\begin{equation}\label{o-a-ansatz-star}
    \dot{\alpha} = -i(\beta-1)\omega\alpha +\frac{\lambda}{2} - \frac{\lambda}{2}\alpha^2 + \frac{\lambda\beta}{2}|\alpha|^2-\frac{\lambda\beta}{2}\alpha^2-D\alpha.
\end{equation}
where $\alpha\equiv\alpha_1$ and we take into account that in our set up $|r|\equiv|\alpha|$. The equation~\eqref{o-a-ansatz-star} is our key result. Let us discuss it carefully.

First of all, one immediately notice that in case of $D=0$, the eq.~\eqref{o-a-ansatz-star} coincides with the equation for order parameter from the paper \cite{Vlasov2015}. It is quite interesting fact because naively setting $D\rightarrow 0$ we obtain nothing meaningful for self-consistency equation and one should carefully analyze uniform expansions of Bessel function, which is quite time-consuming. Second, it is interesting to emphasize that in the paper \cite{Vlasov2015} the authors use W-S ansatz and then take the thermodynamic limit, whereas we use O-A ansatz. It is not a surprise because it is straightforward to show that O-A ansatz coincides with W-S ansatz for continuum limit.

\begin{figure}
    \centering
    \includegraphics[width=\linewidth]{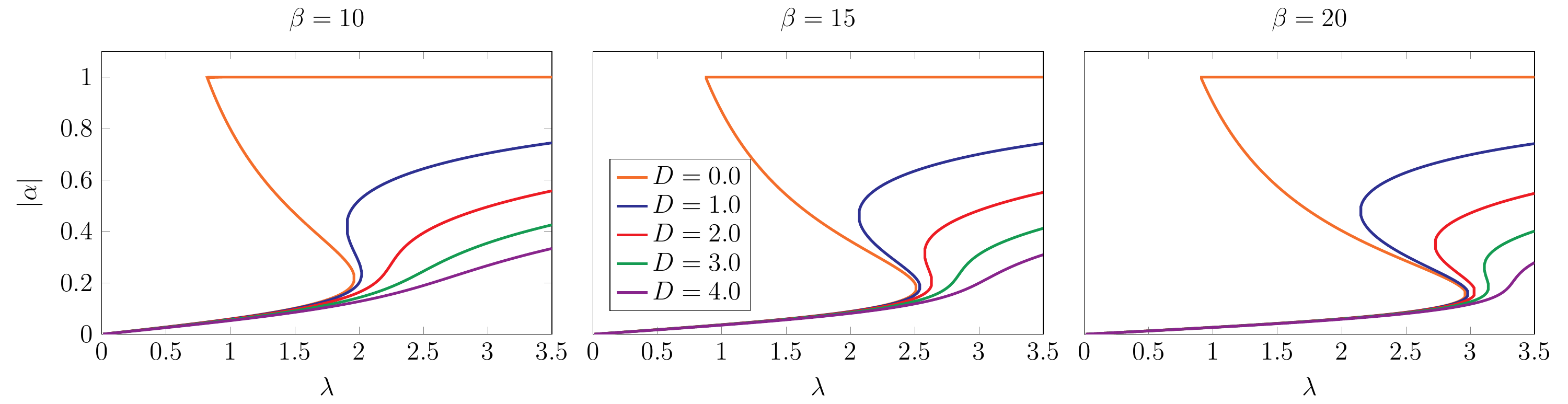}
    \caption{The order parameter ($|r|\equiv|\alpha|$) in thermodynamic limit}
    \label{fig:numerical_fp}
\end{figure}

Using the polar representation of complex number, $\alpha=Ae^{i\phi}$, we obtain the following system of equations for fixed points of~\eqref{o-a-ansatz-star},
\begin{equation}\label{star-noise-fp}
\begin{cases}\displaystyle \dot{A} = \frac{\lambda}{2}\left(1-A^2\right)\cos\phi -DA, \\ \displaystyle\dot{\phi}A=-(\beta-1)\omega A-\frac{\lambda}{2}\sin\phi(1+2\beta)A^2-\frac{\lambda}{2}\sin\phi.\end{cases}
\end{equation}
In compare with the work \cite{Vlasov2015}, the system~\eqref{star-noise-fp} looks quite similar, up to additional term with noise in the first equation. We can exclude the complex number phase from the system~\eqref{star-noise-fp}, which gives the following equation
\begin{equation}\label{r2-eq}
    \frac{4D^2A^2}{\lambda^2(1-A^2)^2}+\frac{4(\beta-1)^2\omega^2A^2}{\lambda^2[(2\beta+1)A^2+1]^2}=1,
\end{equation}
In the limit of zero noise, $D\rightarrow 0$, the first term in the left hand side of eq.~\eqref{r2-eq} becomes negligible in compare with the second term. It gives the simple quadratic equation for $A$ from which follows that the critical coupling coincides with eq.~\eqref{star-cc}. In the opposite limit of the large noise, $D\rightarrow\infty$, the second term becomes negligible in compare with the first term. The equation reduces to
\begin{equation}\label{beta-1-large-D}
    \lambda A^2 + 2DA -\lambda=0,
\end{equation}
which always has roots (the discriminant is always positive) and therefore there is no phase transition (the order parameter as function of $\lambda$ gradually grows with increase of $\lambda$). This interesting behavior hints us that there is a such point in $(\lambda,D)$-space (for fixed $\beta$ and $\omega$), where the hysteresis disappear and the phase transition is continuous. Above this point, there is no phase transition with respect to $\lambda$. This conjecture is proven by the numerical analysis of fixed points of eq.~\eqref{o-a-ansatz-star}, see fig.~\ref{fig:numerical_fp}.
To determine a point where the phase transition becomes continuous, we can set $A^2\equiv z$ and rewrite eq.~\eqref{r2-eq} in terms of new variable. It gives the fourth order polynomial with coefficients that depend on all the model parameters and we should analyze roots of this polynomial on the interval $z\in [0,1]$. We are interested in the situation where such polynomial has exactly two roots (that corresponds to critical points). So, we have to analyze when the discriminant $\Delta(\lambda,\omega,D)$ of the fourth order polynomial is less than zero $\Delta(\lambda,\omega,D)<0$. We should solve this inequality for $\lambda$ and than analyze the case when two critical couplings coincide. It determines a critical value of noise amplitude $D$ with fixed $\omega$ and $\beta$ when the transition becomes continuous. The described program can be done analytically but results are quite complicated to be written explicitly.

\begin{figure}
    \centering
    \includegraphics[width=0.68\linewidth]{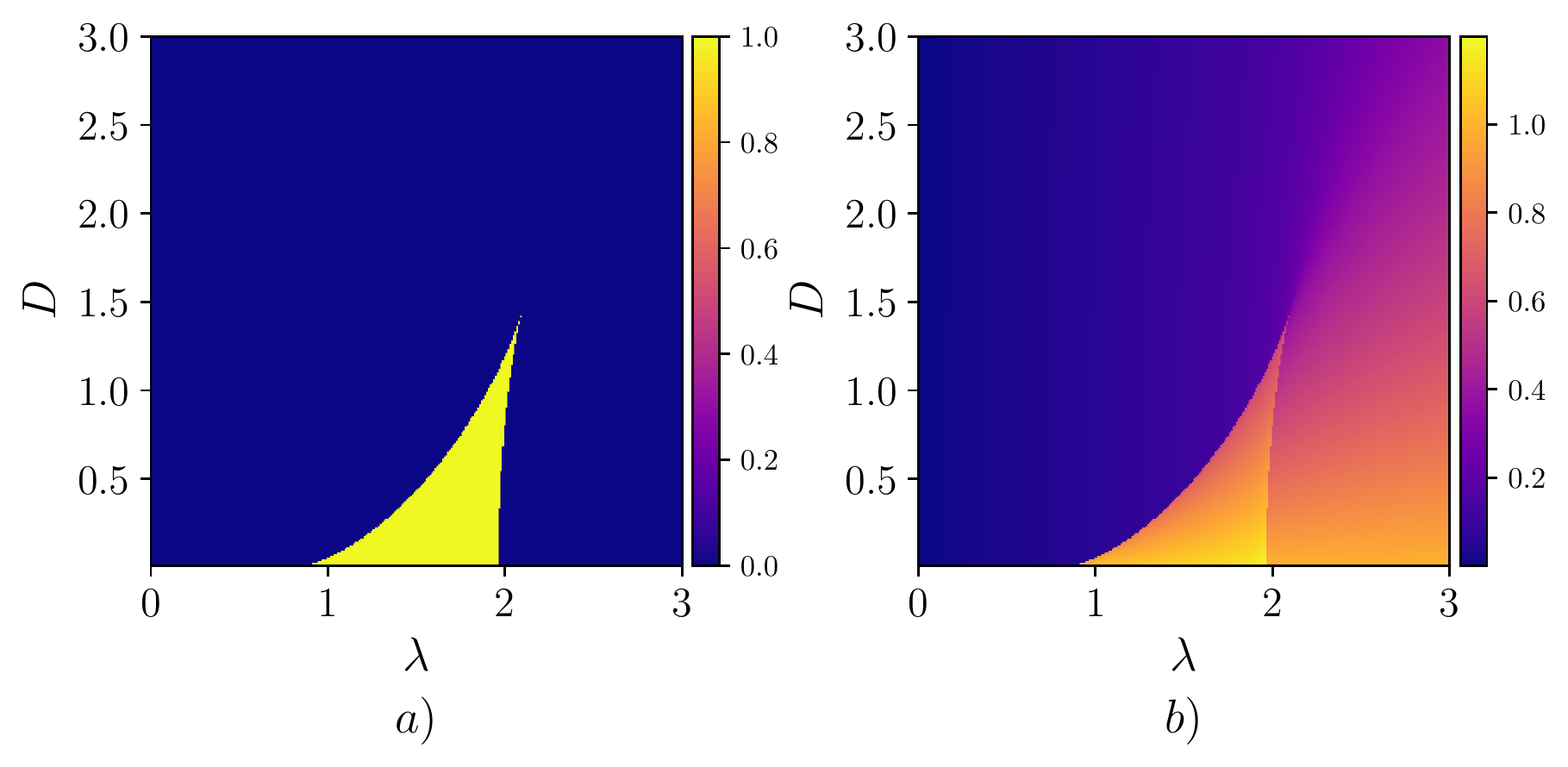}
    \caption{a) Boundaries of the domain in $(\lambda,D)$-space where hysteresis occurs for $\beta=10$ \& $\omega=1$ (color represents does hysteresis exist or not); b) phase diagram in $(\lambda,D)$-space with $\beta=10$, $\omega=1$ (in domain with hysteresis we plot the \emph{sum} of order parameter values)}
    \label{fig:phase_diagramm}
\end{figure}

Hopefully, we can implement the mentioned strategy numerically and obtain quite tractable result. First, we determine the domain in $(\lambda,D)$-space where the discontinuous phase transition occurs. We fix $\beta$ and $\omega$ and then analyze numerically the roots of the polynomial, which gives us the desired domain (see fig.~\ref{fig:phase_diagramm}, left).

\begin{figure}
    \centering
    \includegraphics[width=\linewidth]{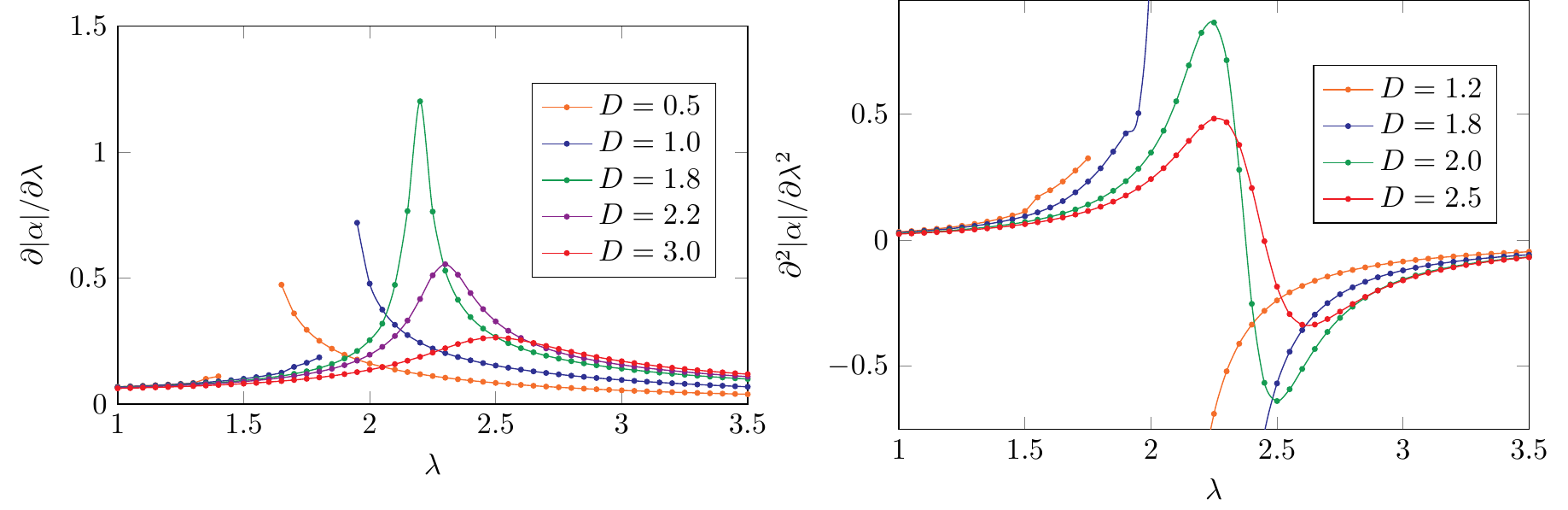}
    \caption{The order parameter ($|r|\equiv|\alpha|$) derivatives in thermodynamic limit}
    \label{fig:op_derivative}
\end{figure}

Second, we can improve our understanding about appeared phase transitions by analyzing numerically the derivatives of $|\alpha|$ as the functions of $\lambda$. We investigate the first and the second derivatives. It is easy to see that for quite small values of $D$ the transition indeed first order, whereas for large enough noise amplitude it is continuous. Than, we also notice that for very large noise the mentioned derivatives are continuous, which means that there is no phase transition (see. fig.~\ref{fig:op_derivative}).
Finally, we can plot the phase diagram in $(\lambda,D)$-space (see fig.~\ref{fig:phase_diagramm}, right). From this diagram, we recognise previously mentioned triangle-shaped region that corresponds to the discontinuous phase transition and the tricritical point. In the next section we provide numerical simulation that reinforces these findings.

\section{Simulations}
\label{sec:Simulations}

In this section we provide simulations of Kuramoto model on star graph with noise in order to support our theoretical results. All the technical details of simulations are provided in the Appendix.

\begin{figure}
    \centering
    \includegraphics[width=\linewidth]{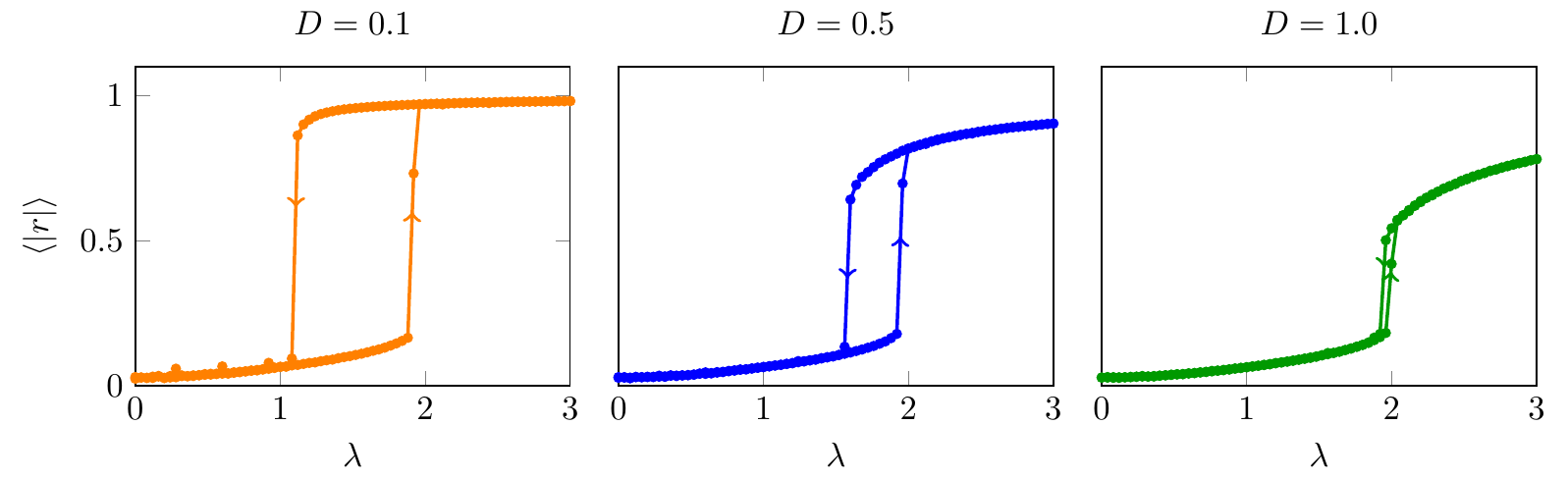}
    \caption{Time-averaged order parameter $\langle |r|\rangle$ for the model with fixed $\beta=10$ and three relatively small values of noise $D$. Arrows indicate evolution (forward or backward)}
    \label{fig:hysteresis_panel_1}
\end{figure}

\begin{figure}
    \centering
    \includegraphics[width=0.68\linewidth]{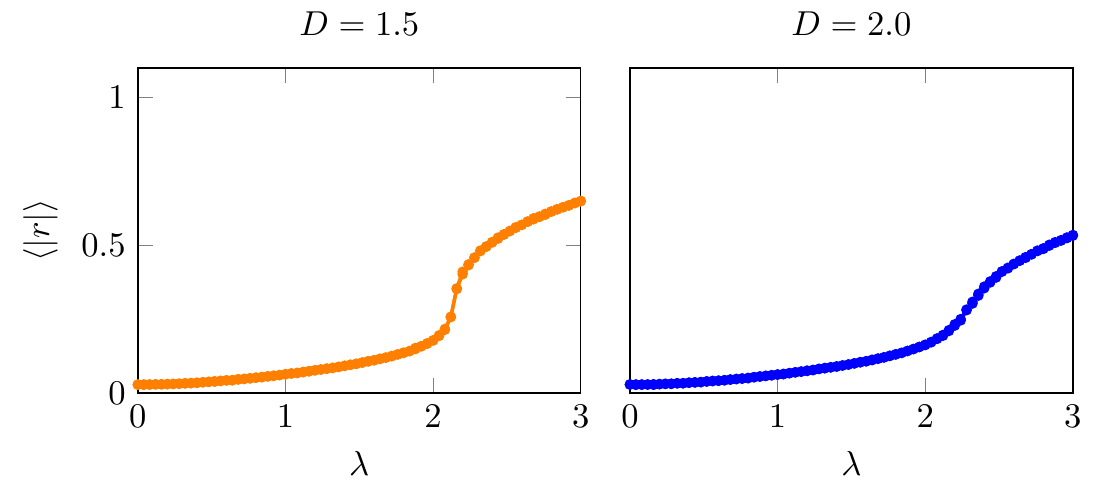}
    \caption{Time-averaged order parameter $\langle |r|\rangle$ for the model with fixed $\beta=10$ and two relatively large values of noise $D$ (forward and backward evolution are shown, there is no hysteresis)}
    \label{fig:hysteresis_panel_2}
\end{figure}

To demonstrate the crossover between phase transitions, we fix the hub frequency enhancement factor $\beta=10$ and vary the amplitude of noise from quite small (fig~\ref{fig:hysteresis_panel_1}) to relatively big values (fig~\ref{fig:hysteresis_panel_2}). From the plots, it is clear to see hysteresis loop becomes smaller and near $D=1.0$ the hysteresis becomes negligible (forward and backward coupling constants become closer to each other) and at vicinity of this point (near $D=1.5$) the phase transition is continuous. For $D=2.0$ the synchronization transition is clearly absent. On the other hand, for small values of $D$ there is the first order phase transition.

\section{Conclusion}
\label{sec:Conclusion}

In this note we have considered the Kuramoto model with noise on two graph types: the well-known case of a complete graph and very similar case of a star graph. We have obtained the self-consistency equation in the closed form. In case of complete graph this closed form is quite simple, whereas in case of star graph the analytical treatment of this self-consistency equation is relatively complicated problem. Nevertheless, we have investigated the phase transition on star graph in the presence of noise both analytically and numerically. Our treatment is consistent with numerical simulations.

Our main result is the crossover between first and second order phase transitions that takes place for a large enough values of noise amplitude $D$. However, we have not succeeded in an analytical derivation of an exact value of $D$ that determines crossover. Nevertheless, we have made numerical calculations which allow us to develop intuition about the mentioned crossover. The limiting cases of vanishing and large noise can be analyzed exactly due to simplicity of the equation for fixed points in these limits. The zero noise case reproduces the well-known result for a star graph, whereas non-zero noise case is analyzed in thermodynamic limit and by numeric simulations, which gives consistent results.

We believe that there is a more profound connection between M\"{o}bius transformation (which is directly related to the low-dimensional dynamics of the Kuramoto model) and considered system on star graph with noise. We address this question for further research and we have a hope that it will be fruitful.

As we have already mention, the noise is common for many biological systems. The Kuramoto model is the good tool to probe synchronization phenomena in biological neuron networks. The star graph can be considered as a motif of these networks. Therefore, we hope that our theoretical consideration can be used in study of a synchronization in neuronal networks.

\section*{Acknowledgements}

I am grateful to Alexander Gorsky for countless fruitful discussions and invaluable comments. I thank Vladimir Arkhipkin for great advice about simulations. I greatly appreciate comments by Mikhail Tamm about the transition crossover and discussion about O-A ansatz with Denis Goldobin. This work is supported by the Brain Program of the IDEAS Research Center.

\section*{Data availability}

The datasets generated during and analysed during the current study are available from the corresponding author on reasonable request.

\appendix

\section{Technical details}
\label{sec:TechnicalDetails}

\subsection{Computation of normalization constant}

The normalization condition gives the following expression for normalization constant $\mathcal{N}$,
\begin{equation}
    \mathcal{N} = \int_{0}^{2\pi}d\phi\,e^{A\phi+B\cos\phi}\int_{\phi}^{\phi+2\pi}d\psi\,e^{-A\psi-B\cos\psi}.
\end{equation}
Introducing the variable $\chi = \psi - \phi$, this can be rewritten in more simple way,
\begin{equation}
    \mathcal{N} = \int_{0}^{2\pi}\int_{0}^{2\pi}d\chi\,d\phi\,\exp\left\lbrace-A\chi + 2B\sin\frac{\chi}{2}\sin\left(\phi+\frac{\chi}{2}\right)\right\rbrace.
\end{equation}
The integration can be performed explicitly. First, integration over $\phi$ gives
\begin{equation}
    \int_{0}^{2\pi}d\phi\,\exp\left\lbrace 2B\sin\frac{\chi}{2}\sin\left(\phi+\frac{\chi}{2}\right)\right\rbrace = 2\pi I_0\left(2B\sin\frac{\chi}{2}\right).
\end{equation}
Second, integral over $\chi$ now becomes
\begin{equation}
    \int_{0}^{2\pi}d\chi\,e^{-A\chi}I_0\left(2B\sin\frac{\chi}{2}\right)=\int_{0}^{\pi}d\chi\,e^{-A\chi}I_0\left(2B\sin\frac{\chi}{2}\right)+\int_{\pi}^{2\pi}d\chi\,e^{-A\chi}I_0\left(2B\sin\frac{\chi}{2}\right).
\end{equation}
For the first integral, let us introduce new variable $y=(\pi-\chi)/2$, so
\begin{equation}
    \int_{0}^{\pi}d\chi\,e^{-A\chi}I_0\left(2B\sin\frac{\chi}{2}\right) = - \int_{\pi/2}^{0}dy\,e^{-\pi A}e^{2Ay}I_0\left(2B\cos y\right),
\end{equation}
and for the second integral we use $y=(\chi-\pi)/2$, which gives
\begin{equation}
    \int_{\pi}^{2\pi}d\chi\,e^{-A\chi}I_0\left(2B\sin\frac{\chi}{2}\right)=\int_{0}^{\pi/2}dy\,e^{-\pi A}e^{-2Ay}I_0\left(2B\cos y\right).
\end{equation}
Combining two obtained integrals, we find
\begin{equation}
    \mathcal{N} = 8\pi e^{-\pi A}\int_{0}^{\pi/2}dy\,\cosh(2Ay)I_0\left(2B\cos y\right).
\end{equation}
Writing down the hyperbolic cosine as trigonometric cosine,
\begin{equation}
    \cosh(2Ay)=\cos(2iAy).
\end{equation}
we obtain the known integral (see 6.681.3 of \cite{Gradshteyn2014}),
\begin{equation}
    \int_{0}^{\pi/2}dy\,\cos(2iAy)I_0(2B\cos y) = \frac{\pi}{2}I_{iA}(B)I_{-iA}(B)
\end{equation}
and finally the constant $\mathcal{N}$ is given by
\begin{equation}
    \boxed{\mathcal{N} = 4\pi^2e^{-\pi A}I_{-iA}(B)I_{iA}(B).}
\end{equation}

\subsection{Computation of order parameter related integrals}

We compute the following integral,
\begin{equation}
    \mathcal{I}=\int_{0}^{2\pi}d\phi\,\cos\phi \,e^{A\phi+B\cos\phi}\int_{\phi}^{\phi+2\pi}d\psi\,e^{-A\psi-B\cos\psi}.
\end{equation}
To proceed the derivation, we use variable $\chi=\psi-\phi$ one more time. The integral becomes
\begin{equation}
    \mathcal{I}=\int_{0}^{2\pi}\int_{0}^{2\pi}d\phi\,d\chi\,\cos\phi\,e^{-A\chi}\exp\left\lbrace 2B\sin\frac{\chi}{2}\sin\left(\frac{\chi}{2}+\phi\right)\right\rbrace.
\end{equation}
Integration over $\phi$ gives us
\begin{equation}
    \int_{0}^{2\pi}d\phi\,\cos\phi\exp\left\lbrace 2B\sin\frac{\chi}{2}\sin\left(\frac{\chi}{2}+\phi\right)\right\rbrace = 2\pi\sin\frac{\chi}{2}I_1\left(2B\sin\frac{\chi}{2}\right).
\end{equation}
Next step is to compute integral over $\chi$. The procedure is quite similar to previous computation with several distinctions. First, after combining two integrals, we write
\begin{equation}
    \cos y\left(e^{2Ay}+e^{-2Ay}\right)=2\cos y\cosh (2Ay) = \cos[(2Ai+1)y]+\cos[(2Ai-1)y].
\end{equation}
So, we should compute two integrals,
\begin{equation}
\begin{gathered}
    J_+ = 4\pi e^{-\pi A}\int_{0}^{\pi/2}dy\,\cos[(2Ai+1)y]I_1(2B\cos y), \\
    J_- = 4\pi e^{-\pi A}\int_{0}^{\pi/2}dy\,\cos[(2Ai-1)y]I_1(2B\cos y).
\end{gathered}
\end{equation}
With help of G-R book, these integrals are
\begin{equation}
\begin{gathered}
    J_{+} = 2\pi^2e^{-\pi A}I_{-iA}(B)I_{iA+1}(B), \\
    J_{-} = 2\pi^2e^{-\pi A}I_{1-iA}(B)I_{iA}(B).
\end{gathered}
\end{equation}
Finally, we write
\begin{equation}
    \boxed{\mathcal{I} = J_{+}+J_{-} = 2\pi^2e^{-\pi A}\left(I_{-iA}(B)I_{iA+1}(B)+I_{1-iA}(B)I_{iA}(B)\right)}
\end{equation}
Next, we compute the integral,
\begin{equation}
    \mathcal{J}=\int_{0}^{2\pi}d\phi\,\sin\phi \,e^{A\phi+B\cos\phi}\int_{\phi}^{\phi+2\pi}d\psi\,e^{-A\psi-B\cos\psi}.
\end{equation}
As usual, introduce $\chi = \psi-\phi$,
\begin{equation}
    \mathcal{J} = \int_{0}^{2\pi}\int_{0}^{2\pi}d\phi\,d\chi \sin\phi e^{-A\chi}\exp\left\lbrace 2B\sin\frac{\chi}{2}\sin\left(\frac{\chi}{2}+\phi\right)\right\rbrace.
\end{equation}
Integration over $\phi$ gives
\begin{equation}
    \int_{0}^{2\pi}d\phi\,\sin\phi\exp\left\lbrace 2B\sin\frac{\chi}{2}\sin\left(\frac{\chi}{2}+\phi\right)\right\rbrace=2\pi\cos\frac{\chi}{2}I_1\left(2B\sin\frac{\chi}{2}\right).
\end{equation}
Integration over $\chi$ can be performed as in previous case. During the computation, we use
\begin{equation}
    \sin y\left(e^{2Ay}-e^{-2Ay}\right)=2\sin y \sinh (2Ay) = i\cos[(2iA+1)y] - i\cos[(2iA-1)y].
\end{equation}
As in previous, we deal with two integrals, $J_{+}$ and $J_{-}$. The final answer is
\begin{equation}
    \boxed{\mathcal{J} = 2i\pi^2e^{-\pi A}\left(I_{-iA}(B)I_{iA+1}(B)-I_{1-iA}(B)I_{iA}(B)\right)}
\end{equation}

\section{Simulation details}
\label{sec:SimulationDetails}

The equations of motion is nothing more than the system of coupled random processes,
\begin{equation}
    \begin{gathered}
        d\Phi = \beta\omega\,dt+\frac{\lambda\beta}{N}\sum_{j=1}^{N}\sin\left(\theta_j(t)-\Phi(t)\right)\,dt,\\
        d\theta_j = \omega\,dt+\lambda\sin(\Phi(t)-\theta_j(t))+dW_t\left(0,\sqrt{2D}\right),
    \end{gathered}
\end{equation}
where $W_t(0,\sqrt{2D})$ is the Wiener process with zero average, $\mu=0$, and volatility $\sigma^2=2D$. The order parameter,
\begin{equation}
    r(t) = \frac{1}{N}\sum_{j=1}^{N}\exp\left(i\theta_j(t)-i\Phi(t)\right)
\end{equation}
is the random process which satisfies Ito's lemma conditions. On plots we deal with
\begin{equation}
    \langle |r|\rangle = \lim\limits_{T\rightarrow\infty}\frac{1}{T}\int_0^{T}dt\,|r(t)|.
\end{equation}
We implement the following simulation strategy (which is common and have been realized in previous works):
\begin{flushleft}
    \noindent\rule{\linewidth}{0.6pt}\\
    {\bf Simulation scheme} \\
    \noindent\rule{\linewidth}{0.6pt} \\
    {\bf set} $\lambda_{\max}$ and $\delta\lambda$\\
    {\bf while} $\lambda \leq\lambda_{\max}$ ($\lambda\geq 0)$\\
    \begin{enumerate}
        \itemsep0em 
        \item choose uniformly distributed random initial conditions $\theta_i(t=0)\in [0,2\pi)$
        \item simulate Ito random process on time interval $[0,T]$ with step $\delta T$, $\delta T\ll T$ and store values $\theta_i(t=T)$
        \item compute time averaged order parameter as follows,
        \begin{equation*}
            \langle |r|\rangle = \frac{\delta T}{T}\sum_{k=1}^{M}\frac{1}{N}\left|\sum_{i=1}^{N}\exp\left(\theta_i(\tau_k)\right)\right|,\quad M=T/\delta T.
        \end{equation*}
        \item use obtained $\theta_i(t=T)$ as initial conditions for the next round of simulations and shift coupling constant by $\delta\lambda$: $\lambda\rightarrow\lambda+\delta\lambda$ ($\lambda\rightarrow\lambda-\delta\lambda$)
    \end{enumerate}
    {\bf end}
\end{flushleft}
The values in parentheses in the description of first and fourth items correspond to the case of backward evolution, i.e. when we start from the synchronized state and decrease coupling constant. 

We have observed that finite size effects are small enough starting from $N=1000$. The finite size effects are following: shift of backward and forward critical coupling (underestimation of couplings in comparison with results from Riccati equation), fluctuations of the order parameter for large couplings. Next, we have observed that $T=300$ and $\delta T=0.01$ are enough for correct simulation. So, unless it is explicitly specified, we set $N=1000$, $T=300$, $\delta T=0.01$. The choice of $\lambda_{\max}$ is not restricted by any numerical issues but due to our computation facility constraints, we choose $\lambda_{\max}=4$ for virtually all simulations and $\delta\lambda=0.04$. It is enough to see synchronization transition for chosen parameters $\beta$ and $D$ (in case of complete graph and large noise values, we use $\lambda_{\max}=10$). For forward evolution we start from $\lambda=0$, whereas for backward evolution we star from $\lambda=\lambda_{\max}$. For each simulation we fix $\omega=1.0$.

All the simulations are performed in Wolfram Mathematica. The method for solution of stochastic differential equation is Euler-Maruyama method.

\bibliography{main.bib}

\end{document}